\documentclass[a4paper,fleqn,usenatbib]{mnras}

\usepackage{color}
\usepackage{booktabs}
\usepackage{graphicx}
\usepackage{amsmath}
\usepackage{amssymb}
\usepackage[bottom]{footmisc}
\usepackage{tabularx}
\usepackage[export]{adjustbox}
\usepackage{multirow}

\usepackage{enumitem}

\newcommand\Bt{\rule[-1.0ex]{0pt}{0pt}}

\newcommand\Tt{\rule{0pt}{2.5ex}}
\usepackage{parskip}

 \title{Kilonova rates from spherical and axisymmetrical models}

	\author{J\'ozsef K\'obori, $^{1}$, Zsolt Bagoly, $^{1}$, Lajos G. Bal\'azs$^{2,3}$
	\\
	$^{1}$Department of Physics of Complex Systems, E\"otv\"os University,
	H-1117 Budapest, P\'azm\'any P. s. 1/A, Hungary\\
	$^{2}$MTA CSFK Konkoly Observatory, Konkoly-Thege M. \'ut 13-17, Budapest, 1121, Hungary\\
	$^{3}$Department of Astronomy, E\"otv\"os University,
	H-1117 Budapest, P\'azm\'any P. s. 1/A, Hungary}
	
\begin{document}

 \maketitle

 \begin{abstract}
 Detecting the thermal emission from double neutron star merger events is a challenging task because of
the quick fading of the observed flux. In order to create an efficient observing strategy
for their observing method it is crucial to know their intrinsic rate. Unfortunately,
the numerous models existing today predict this rate on a vary wide range.
Hence, our goal in this paper is to investigate the effect of different level of approximations 
on the \textit{relative} rate predictions. Also, we study the effect of distinct ejecta mass lay-outs on the light curve.
We find that the ratio of the expected kilonova detections of the spherical to
axisymmetrical models is 6:1 (or 2:1, depending on the input parameter set applied in our work).
Nevertheless, the light curve shape is only slightly affected by the various ejecta alignments.
This means that different ejecta lay-outs can produce
light curves with similar shapes making it a challenging task to infer the structure of the matter outflow.
Thus, we conclude that the uncertainty in the rate predictions arising from the various
ejecta mass distribution models
is negligible compared to the errors present in other input parameters (e.g. binary neutron star merger rate).
In addition, we show that up to moderate redshifts ($z\lesssim 0.2$) the redshift distribution type 
(observed or uniform in volume) does not affect the expected relative rate estimations.

 \end{abstract}


	\section{Introduction}
	\label{sec:introduction}
	The discovery of the GW170717 gravitational wave source \citep{2017PhRvL.119p1101A} in association with the GRB 170817A \citep{abott2017_12}
and its kilonova (hereafter KN) emission \citep{abott2017_13} was
a significant step for multi-messenger astronomy. The observation of the multi-waveband radiation enabled detailed modelling of
the underlying physics in order to constrain the physical parameters shaping the lightcurves.

Having an efficient observing strategy for a sky survey program (e.g. Large Synoptic Survey Telescope (hereafter LSST))
can greatly enhance the probability to detect KNe. There are numerous works which give an estimation for KN rates
for the future sky surveys (e.g. \citet{wollaeger2018, tan2018, pol2018, scolnic2018, mapelli2018, 
chruslinska2018, kruckow2018, jin2018, gomez2018, eldridge2019, cao2018, dominik2013, sun2015, sadowski2008, cowperthwaite2019} (hereafter C19)).
When calculating these rates one has to take into account the short gamma-ray burst (sGRB) rate
determined from observations (e.g. \citet{sun2015, wanderman2015, paul2018, ruffini2018, zhang2018, dietz2011, coward2012, petrillo2013, yonetoku2014}) or from
population synthesis methods (e.g. \citet{bogomazov2007, ziosi2014, belczynski2016, chruslinska2018, saleem2018b, nakar2006}). In addition,
the particular model (e.g. composition, structure) employed in the study can significantly affect the results.
Hence, it is not surprising that the predictions for KN rates cover a wide range of values depending what
is chosen to be the source of a KN (all binary neutron star systems or only those producing short gamma-ray bursts (sGRB)).

Using an arbitrary detetection limit of $m_{\textrm{AB}} = 24.4$ (the $r$-band single-visit 5-$\sigma$ limiting depth of the LSST)
our aim in this work is to determine the \textit{relative} KN rates calculated with spherical and axisymmetrical models.
Moreover, the effect of models with different ejecta alignments on the relative rate is investigated.

Throughout the paper we adopt the standard $\Lambda$CDM cosmological parameters: 
$\mathrm{H}_0 = 67.8 \ \mathrm{km} \  \mathrm{s}^{-1} \ \mathrm{Mpc}^{-1}, \ \Omega_M = 0.308, \  \Omega_{\Lambda} = 0.692$ ~\citep{planck2016}.

	\section{Lightcurve simulations} 
	\label{sec:paramphysics}

	 \label{kn_models}
The light curve simulations were carried out with the MOSFiT software package.\footnote{\url{http://mosfit.readthedocs.io/}}
This is an open-source Python-based code developed by \cite{guillochon2018}, which is able to 
generate synthetic light curves of various types of transient phenomena.

Our two basic models, the spherical and axisymmetrical, are
the ones created by \cite{villar2017b} (hereafter V17) and \cite{perego2017} (hereafter P17), respectively.

\subsection{Spherical (V17) model}
The spherical model, implemented by \cite{villar2017b} and first outlined by \cite{metzger2017}, was motivated by multi-band observations and is composed of two or three ejecta component:
\begin{itemize}
\item "blue" ejecta: most likely it can be identified as a relatively proton-rich (high electron fraction, $Y_{\mathrm{e}}$)
polar dynamical ejecta created by the shock from the collision between the merging neutron stars,
with opacity of $\kappa\approx0.5$ cm$^{2}$ g$^{-1}$ due to the lanthanide-poor material,
\vspace{2mm}
\item "purple" ejecta: probably a delayed neutron-rich ($Y_{\mathrm{e}} < 0.25-0.3$) outflow from the accretion disk formed in the merger,
with opacity of $\kappa\approx3$ cm$^{2}$ g$^{-1}$,
\vspace{2mm}
\item "red" ejecta: essentially the same as the purple component, but composed of lanthanide-rich material,
hence, 	the opacity is larger, $\kappa\approx10$ cm$^{2}$ g$^{-1}$.
\end{itemize}
The emission of the kilonova is powered by the radioactive decay of formerly generated r-process nuclei.
However, only a fraction of this luminosity is converted into the observed flux: the thermal efficiency, $\epsilon_{\mathrm{th}}<1$, can be approximated
with analytical functions derived from numerical simulations \citep{barnes2016}.
The detected spectrum is the sum of the blackbody radiation emerging from each ejecta component.
Fitting the KN lightcurve of the GW170817 event V17 came to the conclusion that the three component model
gives a slightly better fit to the light curve, thus in our work we use this three ejecta model.
\subsection{Axisymmetrical (P17) model}\label{sec:axisym}
\begin{figure}
\includegraphics[width=8cm]{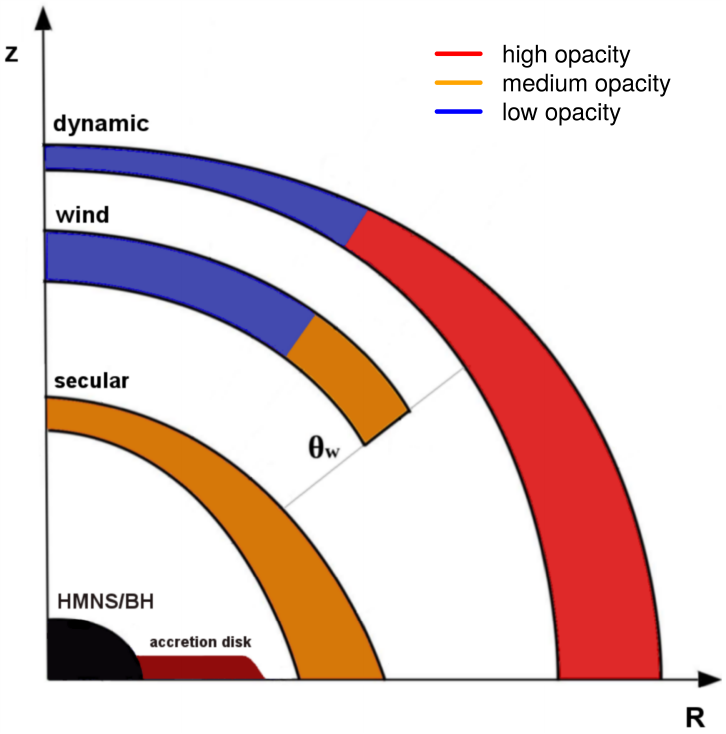}
\caption{The figure shows the structural alignment of the axisymmetrical model created by P17.
The outermost layer is called the dynamic one,
the middle layer 
is labeled as wind and
the innermost layer is the secular one (see Section~\ref{sec:axisym}).
In this model not only the ejecta have their specific mass distribution (see Eq.~\ref{eq:mdist}), but the opacity is changing along the polar axis.
The blue color correspond to a low opacity matter, the orange indicates mild opacities, while the red marks the high opacity matter.
The progenitor object may be a hypermassive neutron star or a black hole,
around which an accretion disk can be formed.
The sketch is based on Figure 2 from P17.
\label{fig:perego_struct}}
\end{figure}

Since the ejecta in the V17 model have spherical arrangement, as a second-order approximation,
we implemented the axisymmetric ejecta structure into this model created by P17.
The main characteristic of this approach is that the mass and opacity of the ejecta have a specific angular distribution
inferred from general-relativistic hydrodynamical simulations. As in the spherical model above P17
uses three ejecta components:
\begin{itemize}
\item dynamical (blue), the mass distribution can be approximated by $F(\theta)=\sin^2 \theta$,
\vspace{2mm}
\item wind (red), most likely polar emission with uniform distribution in mass, $F(\theta)\approx\mathrm{const}$ for $\theta \lesssim \theta_{\mathrm{w}} \approx \pi/3$,
\vspace{2mm}
\item secular (purple), equatorial-dominated flow, $F(\theta)=\sin^2 \theta$,
\end{itemize}
where $F(\theta)$ characterizes the mass distribution according to
\begin{equation}
m_{\mathrm{ej}} = A \sum_{k = 1}^{12} m_{\mathrm{ej, k}} = \sum_{k = 1}^{12} 2 \pi \int_{\theta_k-\Delta\theta/2}^{{\theta_k+\Delta\theta/2}} F(\theta) \sin(\theta) d\theta, \label{eq:mdist}
\end{equation}
where $A$ is the normalization constant.
The radiation mechanism only slightly differs from the one adopted in the spherical V17 model:
the radiation emerging from the two innermost ejecta is reprocessed and emitted again by the outermost envelope.
Apart from the structured ejecta and the reprocessed radiation of the innermost ejecta
the model is identical to the spherical V17 one.
A sketch illustrating the ejecta structure in the axisymmetrical P17 model can be seen in Figure \ref{fig:perego_struct}. 
\subsubsection{Ejecta alignments}
Since the assigning of a particular mass component in the spherical V17 model to a mass component in the axisymmetrical P17 model is not entirely straightforward,
we generated KN samples with various mass distribution layouts (see Table \ref{tab:massdists}). For example,
in the P17 model the dynamic blue ejecta has an equatorial dominated mass distribution 
and the red wind component is enclosed into a cone around the rotational axis, in the literature
it is a common practice to consider the inverted arrangement as well: the blue/dynamical ejecta is the polar component (beside the tidal tail part), while
the red ejecta has the equatorial dominated mass distribution often identified as the wind component.
In addition, some works suggest that both of the blue and red radiation component can
have the same dynamical ejecta source (e.g. \citealt{fernandez2015, wanajo2014}).
Also, some authors propose \citep{arcavi2018ApJ, roberts2011} that because of the uncertainties in the photometric observations it
is impossible to distinguish between the different models.

\subsubsection{Parameter ranges}\label{sec:ranges}
In Table~\ref{tab:inputparams} we show the two distinct parameter sets we used in the simulations.
The first one, denoted as \texttt{parset\textunderscore P17}, was used by P17 to infer the physical parameters of the event AT 2017gfo.
The second one, indicated as \texttt{parset\textunderscore C19}, was used by C19 for making predictions for the LSST and the WFD survey.
In the case of the spherical V17 model we applied only the \texttt{parset\textunderscore C19} set,
however, in the case of the axisymmetrical P17 model we generated light curve samples using
both of the parameters sets (\texttt{parset\textunderscore C19}, \texttt{parset\textunderscore P17}).
All quantities are sampled uniformly from their allowed
intervals.

\begin{table}
\caption{In order to investigate the effect of various mass distribution types on the light curve
samples with different mass distribution lay-outs were generated (see Eq.~\ref{eq:mdist}).}
\label{tab:massdists}
\begin{center}
\begin{tabular}{ccccc}
\toprule
  model        && axisym. P17 && spherical V17\\
\cline{2-4}         \vspace{-3.2mm} \\
\midrule
ejecta  & &$F(\theta)$&&  \\
\midrule
purple &  const. &$\sin^2\theta$& $\sin^2\theta$ &const.\\
red &  $\sin^2\theta$ & const& $\sin^2\theta$ &const.\\
blue & $\sin^2\theta$ & $\sin^2\theta$& const.& const.\\
\midrule
hereafter & P17css & P17scs & P17ssc & V17\\
\bottomrule
\end{tabular}
\end{center}
\end{table}


\begin{table}
\caption{The table shows the ranges for the KN input parameters (see Section~\ref{sec:ranges}). The $M_{\mathrm{disk}}$ indicates the
mass of the disk produced by the massive neutron star before collapsing into a black hole. It has a range of $10^{-2} M_{\odot} < M_{\mathrm{disk}} < 10^{-1} M_{\odot}$.
The mass is in unit of $10^{-2} M_{\odot}$, the speed is in $c$ and the opacity is in $\mathrm{cm}^{-2}$ g$^{-1}$.
All quantities are sampled uniformly from their
allowed intervals.}
\label{tab:inputparams}
\begin{center}
\begin{tabular}{ccc}
\toprule
          & \texttt{parset\textunderscore P17} & \texttt{parset\textunderscore C19} \\
\midrule
Parameter & Range & Range\\
\midrule
$m_{\mathrm{ej, blue}}$ &  0.05 - 5 & 0.5 - 2\\
$m_{\mathrm{ej, purple}}$ &  0.05 $\times M_{\mathrm{disk}}$& 1 - 5\\
$m_{\mathrm{ej, red}}$ &   $<0.03 \times M_{\mathrm{disk}}$ & 0.5 - 2\\
\midrule
$v_{\mathrm{rms,b}}$ & 0.1 - 0.23 & 0.25\\
$v_{\mathrm{rms,p}}$ &  0.017 - 0.04 & 0.15\\
$v_{\mathrm{rms,r}}$ &  0.33 - 0.67 & 0.15\\
\midrule
$\kappa_{\mathrm{b}}$ &  0.5 - 30 & 0.5\\
$\kappa_{\mathrm{p}}$ &  1 - 30 & 3\\
$\kappa_{\mathrm{r}}$ &  0.5 - 1 & 10\\
\bottomrule
\end{tabular}
\end{center}
\end{table}

\renewcommand\tabcolsep{12pt}


	 \label{sec:parameters}
	
	 \subsection{Redshift distribution}\label{redshift}
When simulating the observed redshift distribution of short GRBs from the (a) intrinsic redshift distribution of their progenitor compact merger system
one has to take into account the (b) time delay between the formation of the system and the inspiral.
The former one, (a), more or less traces the star formation history (e.g. \citealt{wanderman2015}),
while the latter one, (b), is usually approximated with empirical functions. Such an approximation was done by \cite{sun2015}, who considered the star formation history from \cite{yuksel2008},
and convolved it with three different time delay distribution models: the power-law, the Gaussian and the lognormal delay model.
However, the power-law delay model has been shown to be inconsistent with the observations (e.g. \citealt{tan2018}).
We used the formula for the lognormal time delay model they give to simulate the redshift distribution for our kilonova samples (Eq. 21 in \citealt{sun2015}).
The maximum distance of events in this work is $D_{\mathrm{max}} \approx 1$ Gpc $(z\approx 0.2)$.

Since the maximum brightness of the KNe simulated in various works makes them detectable only up to $z\lesssim0.1$
it is a common practice to neglect the shape of the redshift distribution function and assume a uniform distribution
in comoving volume in the Universe with a density (e.g. \citealt{wollaeger2018, scolnic2018, saleem2018b}). Thus, we also generated light curve samples for both the V17 and P17scs
models where the redshift
is distributed uniformly in comoving volume between $100$ and 740 Mpc.

\subsection{Binary neutron star merger rate}
There are basically two different approaches when considering double neutron star merger rates: determining the rate from short GRB observations 
or calculating it by employing binary compact object population synthesis methods.
The former method is less reliable, since it is believed to suffer from the uncertainties of various parameters, 
like the beaming factor, redshift, minimum luminosity and the time-delay distribution.
If we assume that \textit{all} of the short GRBs produce KNe then the rate lies between $0.2$ \citep{ghirlanda2016} and $40$ Gpc$^{-3}$ yr$^{-1}$ per $f_{\mathrm{b}}^{-1} = 1 - \cos\theta_{\mathrm{j}}$
\citep{nakar2006}, where
$f_{\mathrm{b}}$ is the beaming factor and $\theta_{\mathrm{j}}$ is the jet half-opening angle. 
Much higher KN rate can be derived if we assume that \textit{all} of the binary neutron star merger events
result in KN, namely as high as 1540 Gpc$^{-3}$ yr$^{-1}$ \citep{abott2017_12}.
The lowest value in this case is 316 Gpc$^{-3}$ yr$^{-1}$, which is taken from \cite{valle2018} and is based on the work of \cite{belczynski2008}.
The minimum and maximum short GRB and binary neutron star merger rates are listed in Table \ref{tab:lurates}.

\begin{table}
\caption{The lowest and highest short GRB rates (Gpc$^{-3}$ yr$^{-1}$ per $f_{\mathrm{b}}^{-1} = 1$) 
and binary neutron star merger rates (Gpc$^{-3}$ yr$^{-1}$) in the literature cover a wide range of values. We take these numbers
as lower and upper limits when calculating expected KN rates.}
\label{tab:lurates}
\scalebox{0.9}{
\begin{tabular}{cccc}
\hline
\multirow{2}{*}{BNS systems}& min & 316  & \Tt \cite{belczynski2008}\\[.7mm]
			   & max & 1540 & \Bt \cite{abott2017_12}\\[.7mm]
    \hline
\multirow{2}{*}{short GRBs}& min & 0.2  & \Tt \cite{ghirlanda2016}\\[.7mm]
			   & max & 40 & \Bt \cite{nakar2006}\\[.7mm]
\hline
\end{tabular}
}
\end{table}

	 \label{sec:redshift_merger}

	\section{Results}
	Having the simulated KN sample we can now determine the expected KN rate and compare these rates in the light of the different models.

\subsection{Spherical vs. axisymmetrical model}
\begin{figure}
\includegraphics[width=9cm]{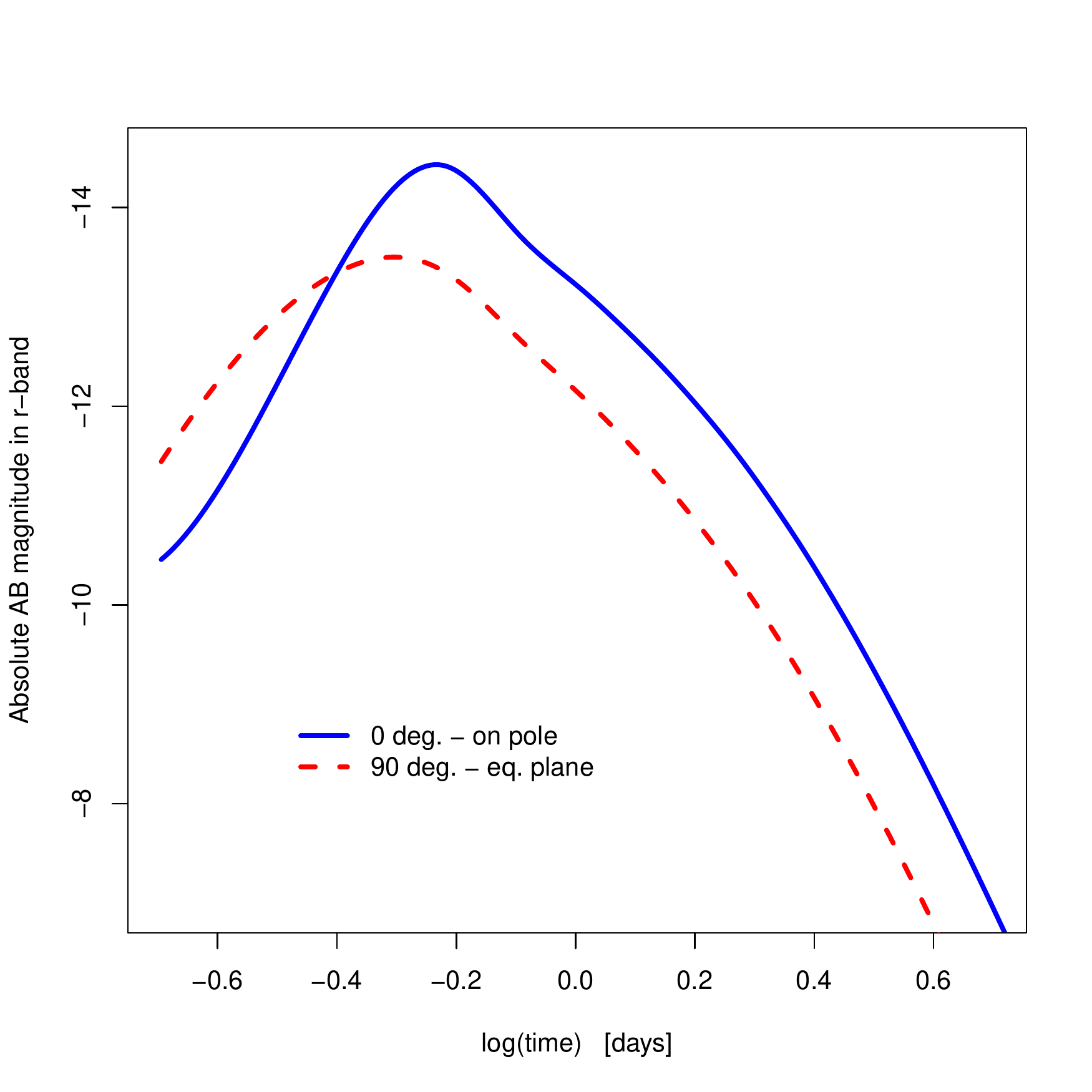}
\caption{The structure of the ejecta can have a significant effect on the light curve shape and the brightness:
radiation coming from an equatorial dominated red ejecta may be obscured if the line of sight is in the equatorial plane of the remnant.
This effect can be seen in the figure: the brightness, if the emission is observed pole-on, is greater since it has 
to traverse only a low-opacity matter. Contrary, an observer in the equatorial plane 
detects the attenuated emission.
\label{fig:obsang}}
\end{figure}

\begin{figure}
\includegraphics[width=9cm]{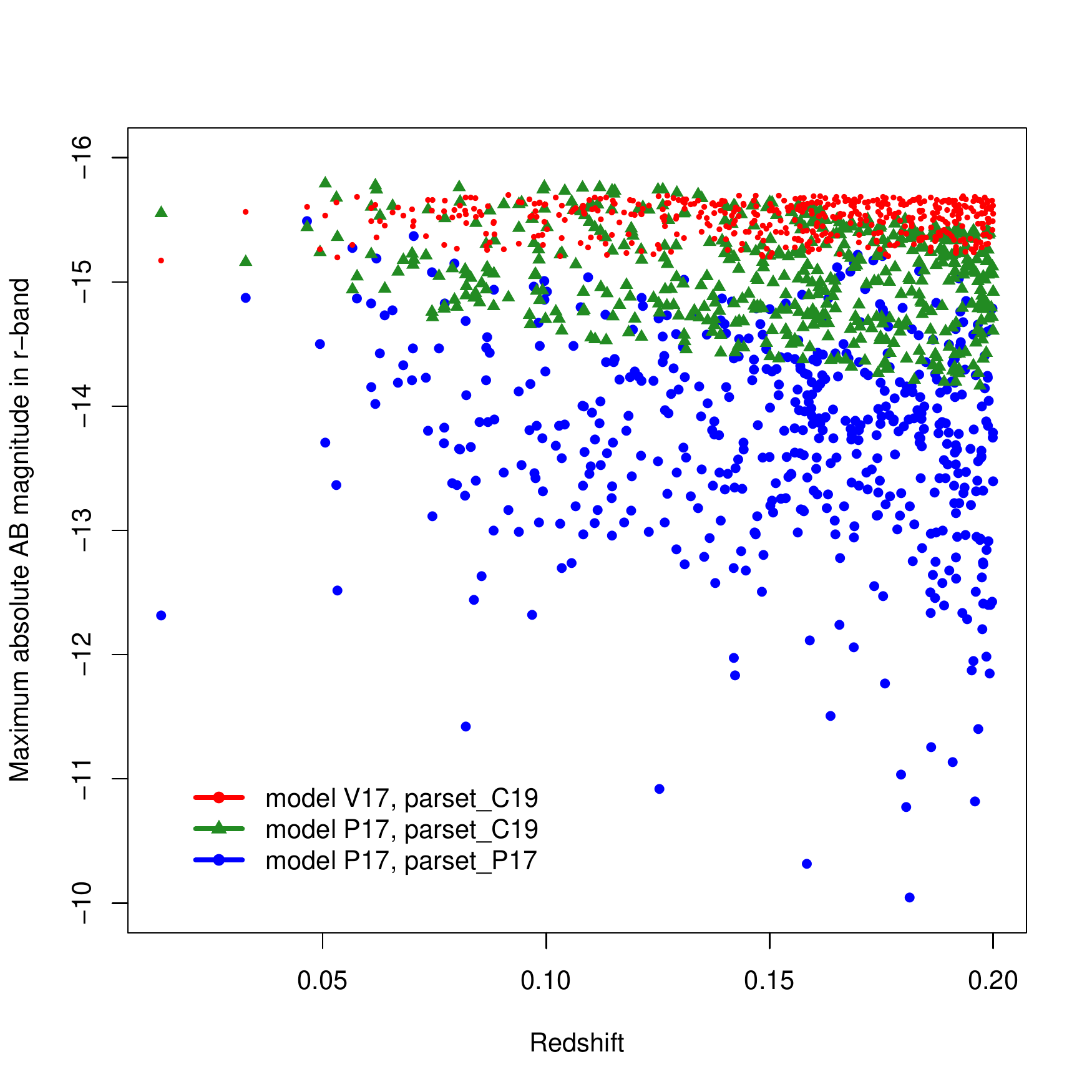}
\caption{The plot shows the maximum $r$-band brightness against the redshift for the distinct samples.
It is observable
that allowing only the ejecta mass to vary in the spherical V17 model (red dots) results in a very narrow maximum brightness distribution in the $r$-band. 
In the case of the axisymmetrical P17 model, the \texttt{parset\textunderscore C19} sample denoted with green triangles,
produces brighter events contrary to the \texttt{parset\textunderscore P17}
sample (blue dots) where the input parameter values are lower compared to the former case (see Section~\ref{sec:ranges}).
\label{fig:V18zmm}}
\end{figure}

\begin{figure}
\includegraphics[width=9cm]{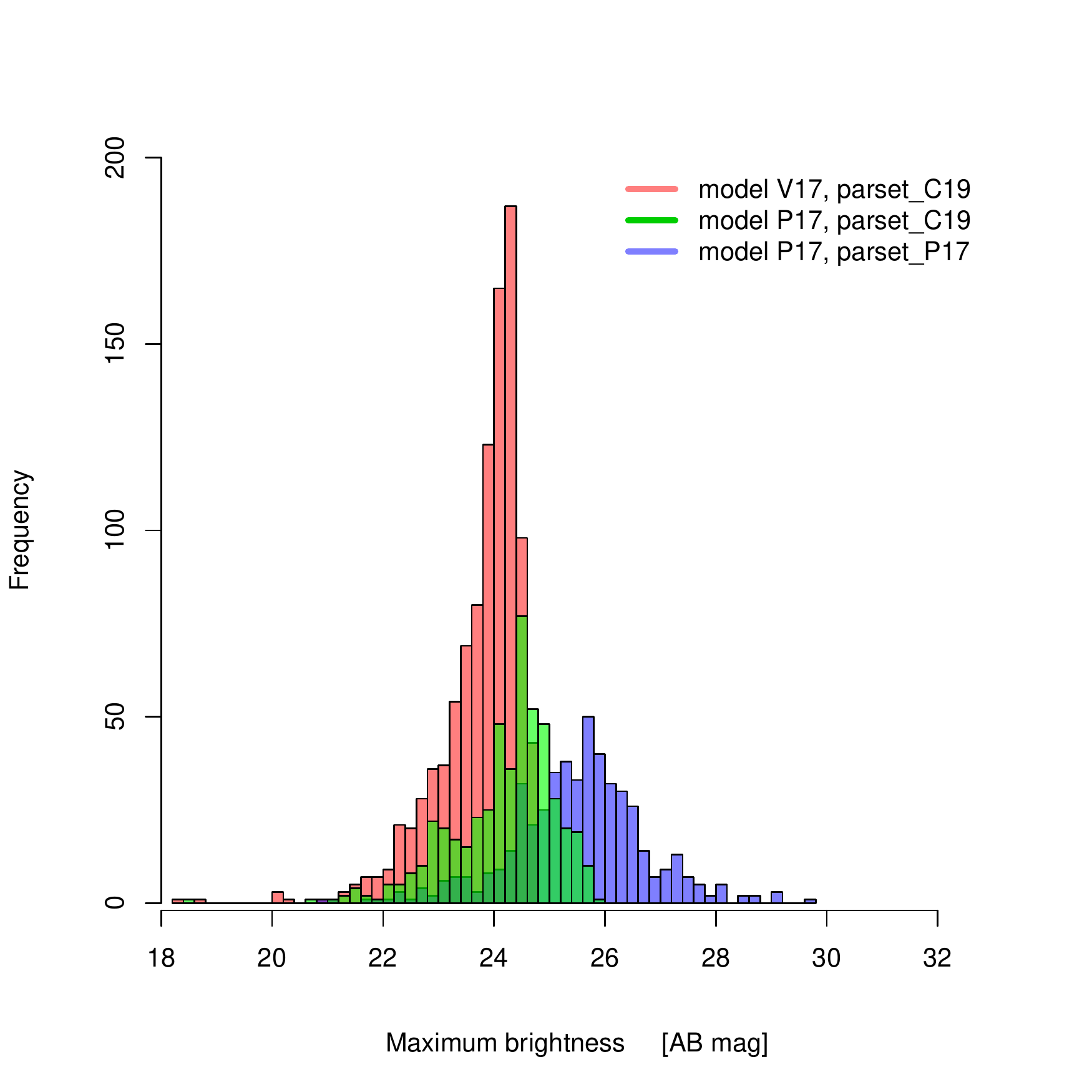}
\caption{This histogram compares the maximum $r$-band brightness of KN samples generated with the spherical and the axisymmetrical model described in Section~\ref{kn_models}.
In the case of the axisymmetrical P17 model the two different parameter sets applied results in different 
maximum brightness distributions: the green histogram indicates the sample generated with \texttt{parset\textunderscore C19},
while the blue histogram shows the sample simulated with \texttt{parset\textunderscore P17} (the original parameter values from \protect \cite{perego2017}).
\label{fig:all_y}}
\end{figure}

Although, as pointed out in V17 their spherical model captures the main characteristics of a KN, there are two main drawbacks.
First, the spherical structure of the ejecta does not reflect the angular alignment of the progenitor stars rotational axis. The result of this is that when
the observer faces the polar or equatorial region of the progenitor the radiation coming from an otherwise obscured ejecta component is
thus overestimated, or similarly, underestimated. This can lead to false predictions for the underlying physical parameters, e.g. ejecta mass.
The difference is clearly visible on the $r$-band light curve (see Figure ~\ref{fig:obsang}).
Second, allowing only the mass to vary (see Table~\ref{tab:inputparams}) strongly narrows down the potential progenitor system's physical parameter space and hence
the width of the resulting maximum brightness distribution. This can be seen in Figure ~\ref{fig:V18zmm} and in Figure \ref{fig:all_y}: the former one shows
the maximum brightness against the redshift of the V17 and P17scs samples,
while the latter one displays the maximum brightness histograms. 
As it can be observed in Figure~\ref{fig:V18zmm} the two different parameter sets (Table~\ref{tab:inputparams}) used for the
axisymmetrical P17 model produces slightly distinct maximum light curve brightness distributions.
While the parameters from P17 (\texttt{parset\textunderscore P17}) can generate events (marked with blue in Figure~\ref{fig:V18zmm}) as bright as the spherical V17 model
only up to $z\sim0.1$, simulations with the higher input values adopted by C19 (\texttt{parset\textunderscore C19}) can clearly reproduce
the bright events (indicated with green) observed also with the spherical V17 model (red in Figure~\ref{fig:V18zmm}) up to $z\simeq0.2$.
The lower brightness of events with higher redshift is a consequence of three factors which are related to each other: first,
the shifting of the spectrum out of the observed band, second, the different contributions of the blue, purple and red ejecta,
and third, the observing angle.

\subsection{Mass distribution lay-outs}
\begin{figure}
\includegraphics[width=8.5cm]{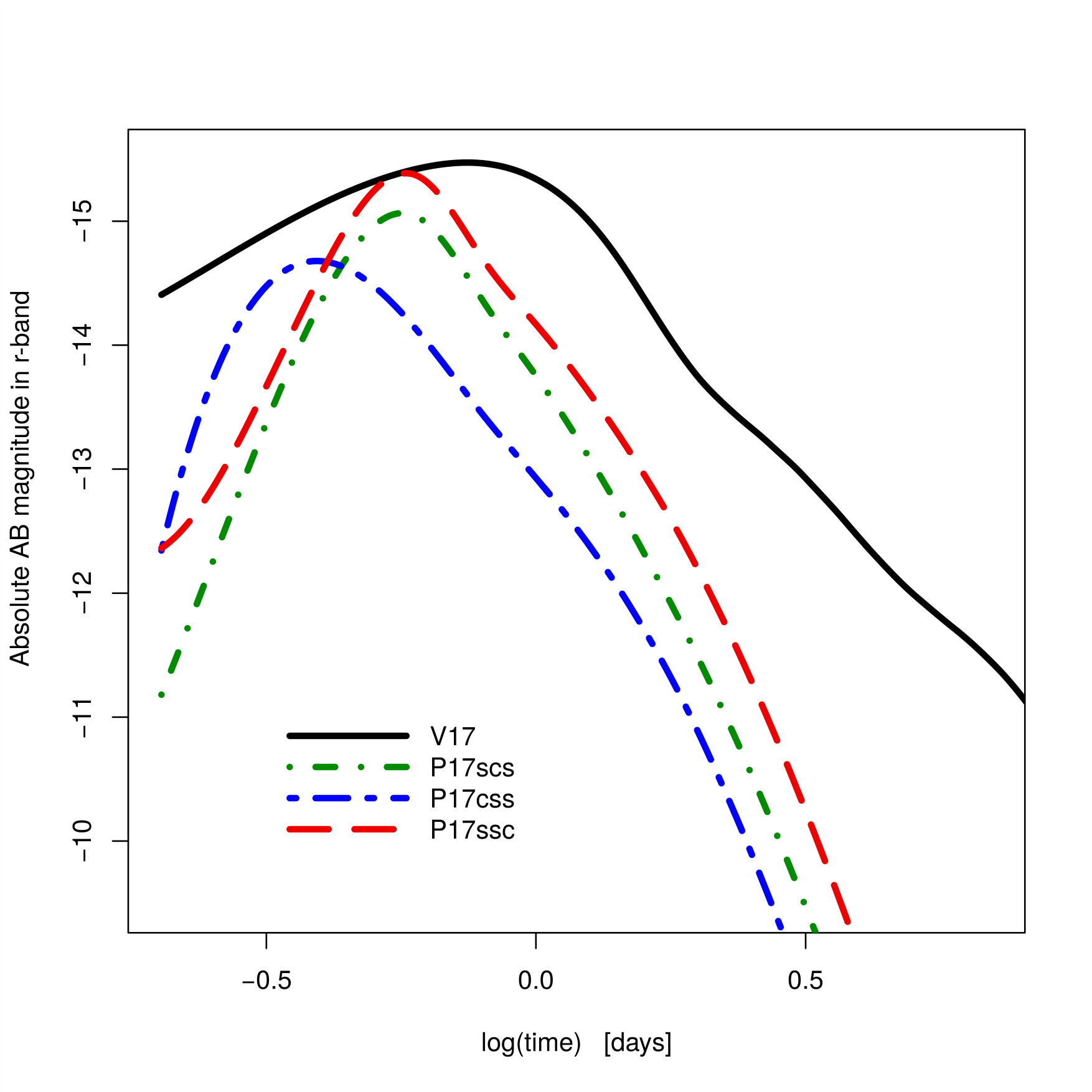}
\caption{A comparison between the models used in this paper shows that the different mass distribution of the ejecta components
can have significant effect on the maximum brightness. The P17css model is $\sim1$ magnitude dimmer than the
P17ssc.
All of the light curves were generated in the $r$-band with the same input parameter values:
$\kappa_d = 0.5$ cm$^{2}$/g,  $\kappa_s = 3$ cm$^{2}$/g,  $\kappa_r = 10$ cm$^{2}$/g,
$m_{\mathrm{ej},b} = 0.011$ M$_{\odot}$, $m_{\mathrm{ej},p} = 0.017$ M$_{\odot}$, $m_{\mathrm{ej},r} = 0.012$ M$_{\odot}$,
$v_{\mathrm{ej},b} = 0.25c$, $v_{\mathrm{ej},p} = 0.15c$, $v_{\mathrm{ej},r} = 0.15c$,
$T_b = 800$ K, $T_p = 1250$ K, $T_r = 3800$ K.
\label{fig:all_peregos_r}}
\end{figure}

As we mentioned it earlier in Section \ref{kn_models}, the KN models commonly used in the literature differ not only in the structural composition (spherical vs. axisymmetrical), 
but in the different ejecta mass distribution types. 
Theoreticaly, this can lead to light curves showing different shapes.
To investigate this, using the axisymmetrical P17 model, we generated KN light curves where we permutted the mass distribution type of the ejecta components, 
changing the opacity accordingly.
Our simulations suggest that the various ejecta configurations do not effect significantly the light curve shapes.
This result is observable in Figure~\ref{fig:all_peregos_r}, where the light curves with different colors and linetypes 
correspond to different models: the thick grey line shows the spherical V17 case, the dashed red line
denotes the P17ssc model, the green dashed-dotted line marks the P17scs case and the blue two-dashed line indicates the
P17css. 
Nevertheless, examining the maximum brightness of the light curves it can be seen that the difference in the maximum brightness can
reach up to $\sim1$ magnitude: the P17css model produces dimmer events than the P17ssc model. 
The steeper rising and decreasing of the observed flux along with the earlier peak time
might be the consquence of fixing the opacity which should be time and wavelength dependent.
However, since we are interested only in the maximum brightness of the events (the ratio of the detectable events simulated with the spherical V17 and axisymmetrical P17
models depends only on the maximum brightness),
we believe that the former shortcomings do not affect our results.
Improving our model with time and wavelength dependent opacity is the scope of a future work.


\subsection{Expected kilonova rate ratios}\label{lsst_rate}

The main problem when calculating the expected KN rate for a specific sky survey program, in our opinion,
is that the uncertainties of the input parameters (e.g. binary neutron star merger rate, mass component distribution)
are relatively high making the estimation unreliable even in the case when the observing cadence is
simulated properly. Because of this we think it is more
appropriate to calculate the \textit{relative} predicted rate of KNe simulated with different approximations.
However, we still need an arbitrary detection limit in order to be able to make an estimation.
For this purpose we use the detection threshold of $m_{\textrm{AB}} = 24.4$, which is the $r$-band 
single-visit 5-$\sigma$ limiting depth of the LSST \citep{LSSToverview}.

Now we can compare the expected rates from different approximations  based on the following aspects:
\begin{enumerate}[label=\alph*), leftmargin=15pt]
\item spherical V17 vs. axisymmetrical (P17) model -
we find that the ratio of the rates with the spherical to the axisymmetrical model is $6:1$.
However, if the higher input parameter values are adopted from C19 the
ratio becomes smaller, $2:1$.

\vspace{1mm}
\item observed vs. uniform in comoving volume redshift distribution - 
considering the spherical V17 model, the ratio of detectable kilonovae simulated with the uniform redshift distribution 
in volume up to $z\lesssim0.15$ to kilonovae simulated with the observed redshift distribution is $\sim1$. 
This is also true for the axisymmetrical (P17) model.
\vspace{1mm}
\item different mass distribution lay-outs - the relative rates of events generated with different mass distribution
lay-outs cover a relatively wide range:\\
rate(P17scs)/rate(P17ssc) $\approx$ 0.6,\\
rate(P17scs)/rate(P17css) $\approx$ 2, \\ 
rate(P17ssc)/rate(P17css) $\approx$ 3.\\
These ratios can be explained by the fact that the discrepancy of the maximum brightness of the events simulated with the different models
can be as large as $\sim$1 magnitudes.
However, the light curves have similar shapes for the distinct models. This can be observed in Figure~\ref{fig:all_peregos_r}.
\end{enumerate}

\begin{table*}
\caption{The table shows the estimated kilonova relative detection rate for LSST up to $z = 0.2$.
The second column (rate(V17)/rate(P17x)) shows the ratio of expected kilonova events of the spherical V17 model to the particular axisymmetrical P17 model
listed in the first column, e.g. there are $\sim6$ times more detectable kilonova with the V17 model than with the P17scs model.
The high values indicate significantly different expected rates for the distinct models.
}
\label{tab:relative_rate}
\begin{center}
\begin{tabular}{ccc}
\midrule
  P17x & rate(V17)/rate(P17x)\\ 
\midrule
P17scs + \texttt{parset\textunderscore P17} & $\sim 6$ \\[1.5mm] 
P17ssc + \texttt{parset\textunderscore P17} & $\sim 4$ \\[1.5mm] 
P17css + \texttt{parset\textunderscore P17} & $\sim 13$  \\[1.5mm] 
P17scs + \texttt{parset\textunderscore C19} & $\sim 2$ \\[1.5mm] 
\bottomrule
\end{tabular}
\end{center}
\end{table*}

	\label{sec:results}

	\section{Conclusions}
	\label{sec:conclusion}
	In this paper we simulated KN events following double neutron star merger events
(along with short gamma-ray bursts).
In order to investigate the effect of the structural compositon of
the underlying physical model on the relative kilonova rate predictions we created samples with spherical and axisymmetrical models.
In addition, different ejecta composition lay-outs were explored to analyze their impact on
the light curve shape. Since the kilonova rate predictions suffer from large errors because of the
uncertainties of the input parameters we calculated only the \textit{relative} rates of the different models.
Our results show that 
\begin{itemize}
\item the ratio of the rates of the events calculated with the spherical and axisymmetrical ejecta structure
can be as small as 2:1. This means that the structural alignment of the kilonova ejecta
is not a significant factor when determining the expected kilonova rates,
\vspace{2mm}
\item changing the distribution type of the distinct mass components does not alter significantly the light curve shape.
Nevertheless, the difference in maximum brightness of the light curves produced with different lay-outs can reach up to $\sim$1 magnitudes,
\vspace{2mm}
\item our simulations support the common assumption that a uniform in comoving volume redshift distribution can be used
when simulating KN events because of their proximity:
the number of observable KNe (up to an arbitrary detetction limit) generated with the spherical V17 and axisymmetrical
P17 model is the same with uniform and observed redshift distribution.
\end{itemize}

Based on our work it can be seen that when calculating the expected
kilonova rates the chosen physical model (spherical, axisymmetrical)
does not affect significantly the results.
In contrast, the uncertainties in the model input parameters, such as the
binary neutron star merger rate (where the predictions cover $2\sim3$ orders of magnitude), play a much more important role
in determining the expected number of events.
Thus, in the light of the above results we think that the ultimate test for the kilonova rates will be
made by the upcoming sky surveys.



\section*{Acknowledgement}
We kindly thank the anonymous referee for the valuable suggestions which significantly improved the quality of this paper.
We also thank the helpful discussions with Ashley Villar, James Guillochon  and P\'eter Veres.

 \bibliographystyle{mnras}
 \bibliography{lsst}

 \end{document}